\documentclass[twocolumn,showpacs,showkeys,preprintnumbers,amsmath,amssymb]{revtex4-1}
\usepackage{bbm}
\usepackage{amsfonts}

\usepackage{epsfig}
\usepackage{graphicx}
\usepackage{amssymb}   
\usepackage{dcolumn}
\usepackage{bm}
\usepackage[colorlinks,citecolor=blue, linkcolor=blue,hyperindex,CJKbookmarks]{hyperref}
\begin{document}

\vspace{2.5cm}
\title{Robust Spin Squeezing Preservation in Photonic Crystal Cavities }
\author{Wo-Jun Zhong$^{1}$}
\author{Yan-Ling Li$^{2}$}
\author{Xing Xiao$^{1}$}
\altaffiliation{xiaoxing1121@gmail.com}
\author{Ying-Mao Xie$^{1}$}
\altaffiliation{xieyingmao@126.com}
\affiliation{$^{1}$College of Physics and Electronic Information,
Gannan Normal University, Ganzhou Jiangxi 341000, China\\
$^{2}$School of Information Engineering, Jiangxi University of Science and Technology, Ganzhou, Jiangxi 341000, China}

\begin{abstract}
 We show that the robust spin squeezing preservation can be achieved by utilizing detuning modification for an ensemble of $N$ separate two-level atoms embedded in photonic crystal cavities (PCC). In particular, we explore the different dynamical behaviors of spin squeezing between isotropic and anisotropic PCC cases when the atomic frequency is inside the band gap. In both cases, it is shown that the robust preservation of spin squeezing is completely determined by the formation of bound states. Intriguingly, we find that unlike the isotropic case where steady-state spin squeezing varies smoothly when the atomic frequency moves from the inside to the outside band edge, a sudden transition occurs for the anisotropic case. 
The present results may be of direct importance for, e.g., quantum metrology in open quantum systems.

\end{abstract}


\maketitle

\section{Introduction}
\label{intro}
Uncertainty relation states that the uncertainty product of any two incompatible observables has a minimum, but it is possible to reduce the uncertainty of one desired observable below the standard quantum limit at the expense of increasing the complementary one. For spin or angular momentum systems, it is often named as spin squeezing if the variance of one angular momentum component, e.g., $(\Delta J_{x})^2$ or  $(\Delta J_{y})^2$ is smaller than $\langle J_{z}\rangle/2$ \cite{ueda93,wineland92}. In recent years, spin squeezing has attracted considerable attention and has been studied both theoretically and experimentally \cite{kuzmich00,ma11,bennett13,vita14}, since it has potential applications in entanglement detection \cite{sorensen01,toth07,guhne09}, quantum information processing \cite{nielsen00} and high-precision measurement, such as Ramsey spectroscopy \cite{wineland94}, atom clocks \cite{meiser08}, gravitational-wave interferometers \cite{goda08} and quantum metrology \cite{riedel10,sewell12,mussel14}.

Unfortunately, any real quantum system inevitably interacts with its surrounding
environment \cite{breuer02}. The environment-induced decoherence effects on spin squeezing have attracted considerable attentions. Numerous researchers have indeed demonstrated that spin squeezing is fragile and easily broken by environmental noise \cite{ulam01,andre02,li08,wang10,sinatra12}. In particular, similar to the discovery of entanglement sudden death (ESD) \cite{yu04}, the sudden death of spin squeezing occurs due to decoherence \cite{wang10}, which would be the most limiting factor for the applications of spin squeezing in realistic tasks. Significantly, non-Markovian effect has been shown potential advantages for enhancing quantum correlations \cite{bellomo08,xiao10a,maniscalco08,xiao10b,matsu11,chin12,huelga12,yang13}, parameter-estimation precision \cite{li15} as well as spin squeezing \cite{yin12} in open quantum systems.

Although the spin squeezing could be partially recovered by non-Markovian effect after a sudden vanishing, it still decays with time and disappears very fast \cite{yin12}. Therefore, it is crucial to pursue some strategies that can effectively protect the spin squeezing from decoherence for a long time. In this paper, we show that the robust preservation of initially prepared spin squeezing states could be achieved with $N$ noninteracting qubits (specifically, two-level atoms) locally embedded in $N$ uncorrelated and identical PCC. Since there is no interaction between the any two pairs of ``qubit+PCC'', the exchange symmetry between qubits still holds during the dynamical evolution. Thus, the spin squeezing parameter of $N$ qubits can be calculated by two-qubit dynamics of local expectations and correlations \cite{wang10,yin12}. With the analytical expression of spin squeezing parameter in mind, we show that, the high spin squeezing could be preserved in both isotropic and anisotropic PCC in the long time limit. The underlying mechanism is due to the permanent existence of a localized field, and hence the photon-atom bound dressed states are formed which lead to a fractionalized steady-state spin squeezing.

This paper is organized as follows. In section \ref{sec:2}, we review the fundamental concept of spin squeezing and introduce a spin squeezing parameter which is extensively studied in both theory and experiment. In section \ref{sec:3}, the exact dynamics of a two-level atom trapped in isotropic and anisotropic PCC are examined. Moreover, we find that these dynamical procedure could be characterized by the Kraus operators of non-Markovian amplitude damping noise. In the section \ref{sec:4}., we show that the spin squeezing could be drastically preserved when the atomic frequency is inside the band gap. The different behaviors of spin squeezing between isotropic and anisotropic PCC are studied in detail. Finally, we summarize our conclusions in section \ref{sec:5}.

\section{Spin squeezing parameters}
\label{sec:2}
Let us consider an ensemble of $N$ spin-1/2 particles and define the collective spin operators as
\begin{equation}
J_{\alpha}\equiv\frac{1}{2}\sum_{m=1}^{N}\sigma_{\alpha}^{(m)},
\label{eq1}
\end{equation}
where $\alpha\in\{x,y,z\}$ and $\sigma_{\alpha}^{(m)}$ are the Pauli operators for the $m$-th particle.
According to the Heisenberg uncertainty relation, the variances of the collective spin components are bounded by the following equation:
\begin{equation}
(\Delta J_{x})^2(\Delta J_{y})^2\geq\frac{1}{4}\langle J_{z}\rangle^2,
\label{eq2}
\end{equation}
with $(\Delta J_{\alpha})^2\equiv\langle J_{\alpha}^2\rangle-\langle J_{\alpha}\rangle^2$.
For a spin squeezed state, the $(\Delta J_{x})^2$ or $(\Delta J_{y})^2$ is smaller than the standard quantum limit $\langle J_{z}\rangle/2$. In this paper, we consider the initial state of the $N$ qubits is
generated in the one-axis twisted spin squeezed state
\begin{equation}
|\Psi(0)\rangle=\exp^{-i\theta J_{x}^2/2}|g\rangle^{\otimes N},
\label{eq3}
\end{equation}
where $|g\rangle$ and $|e\rangle$ denote the ground and excited states. To quantify the degree of useful spin squeezing, there are various measures of spin squeezing related to various inequality criteria \cite{ma11}. Here, we focus on the spin squeezing parameter $\xi_{R}^2$  which is proposed by Wineland \emph{et al} \cite{wineland94}. Note that $\xi_{R}^2$ is substantially connected to the improvement of the sensitivity of Ramsey spectroscopy, and thus is attractive for experimental implementation. The squeezing parameter $\xi_{R}^2$ is defined as the ratio of the phase sensitivity of a general state versus the coherent spin state
\begin{equation}
\xi_{R}^2=\frac{N(\Delta\boldsymbol{J}_{\bot})^2_{\min}}{\langle\vec{\boldsymbol{J}}\rangle^2},
\label{eq4}
\end{equation}
where the minimization of $(\Delta\boldsymbol{J}_{\bot})^2$ is over all the directions that are perpendicular to the mean spin direction $\langle\vec{\boldsymbol{J}}\rangle/\langle\vec{\boldsymbol{J}}^2\rangle$.

Note that the spin squeezing parameter $\xi_{R}^2$ could be expressed as the function of local expectations and correlations due to the exchange symmetry of the one-axis twisted state \cite{wang10,yin12}. Thus, $\xi_{R}^2$ could be written as follows
 \begin{equation}
\xi_{R}^2=\frac{1+2(N-1)\left[\langle\sigma_{+}^{(1)}\sigma_{-}^{(2)}\rangle-|\langle\sigma_{-}^{(1)}
\sigma_{-}^{(2)}\rangle|\right]}{\langle\sigma_{z}^{(1)}\rangle^2},
\label{eq5}
\end{equation}
where $\sigma_{+}=|e\rangle\langle g|$ and $\sigma_{z}$ are the system raising and inversion operators, respectively.
For the initial state of (\ref{eq3}), these local expectations and correlations are calculated as
\begin{eqnarray}
&\langle\sigma_{z}^{(1)}\rangle_{0}=-\cos^{N-1}(\frac{\theta}{2}),\\
\label{eq6}
&\langle\sigma_{+}^{(1)}\sigma_{-}^{(2)}\rangle_{0}=\frac{1-\cos^{N-2}\theta}{8},\\
\label{eq7}
&\langle\sigma_{-}^{(1)}\sigma_{-}^{(2)}\rangle_{0}=-\frac{1-\cos^{N-2}\theta}{8}-\frac{i\sin(\frac{\theta}{2})\cos^{N-2}(\frac{\theta}{2})}{2}.
\label{eq8}
\end{eqnarray}

In order to exhibit the preservation of spin squeezing more conveniently, we re-express the spin squeezing parameter as follows
\begin{equation}
\zeta_{R}^2\equiv\max\{0,1-\xi_{R}^2\}.
\label{eq9}
\end{equation}
Note that  $0<\zeta_{R}^2\leqslant1$ for spin squeezed states and $\zeta_{R}^2=0$ for coherent spin states.

\section{The Physical Model}
\label{sec:3}
Our physical model contains $N$ independent and identical subsystems in which each qubit is embedded in a PCC. Since there are no interactions at all between the subsystems, the whole Hamiltonian can be described via the sum of $N$ independent qubit plus PCC Hamiltonians. In the rotating wave approximation, the Hamiltonian of each subsystem ``qubit+PCC'' is  \cite{john94}
\begin{equation}
H=\hbar\omega_{0}\sigma_{+}\sigma_{-}+\sum_{k}\hbar\omega_{k}a_{k}^{\dagger}a_{k}+i\hbar(\sigma_{-}B^{\dagger}-\sigma_{+}B),
\label{eq10}
\end{equation}
where $B=\sum g_{k}a_{k}$ and $a_{k}^{\dagger}(a_{k})$ is the creation (annihilation) operator of $k$-th mode of the PCC.
$g_{k}=(\omega_{0}d/\hbar)\sqrt{\hbar/2\varepsilon_{0}\omega_{k}V_{0}}\textbf{e}_{k}\cdot\textbf{u}_{d}$ is the strength of coupling between the qubit and the $k$-th mode, where $d$ and $\textbf{u}_{d}$ are, respectively, the magnitude and unit vector of the atomic dipole moment. $V_{0}$ is the quantization volume, and $\varepsilon_{0}$ is the Coulomb constant.

We assume that the PCC is initially in the vacuum state and the atom is prepared in the excited state. Since only one excitation is involved in this subsystem and the total excitation number $N=\sigma_{+}\sigma_{-}+\sum_{k}a_{k}^{\dagger}a_{k}$ of equation (\ref{eq10}) is conserved, then the state vector of the subsystem at an arbitrary time $t$ has the form
\begin{equation}
|\psi(t)\rangle=q(t)e^{-i\omega_{0}t}|e,\textbf{0}_{k}\rangle+\sum_{k}q_{k}(t)e^{-i\omega_{k}t}|g,\textbf{1}_{k}\rangle.
\label{eq11}
\end{equation}
The state vector $|\textbf{0}_{k}\rangle$ denotes no excitation existing in any mode of the PCC, and $|\textbf{1}_{k}\rangle$ represents one excitation in the $k$-th mode. According to the time-dependent
Schr\"{o}dinger equation and in the interaction picture, we can obtain the following integrodifferential equation of $q(t)$
\begin{equation}
\dot{q}(t)+\int_{0}^{t}G(t-\tau)q(t)d\tau=0,
\label{eq12}
\end{equation}
where
\begin{equation}
G(t-\tau)=\sum_{k}g_{k}^2\exp[-i(\omega_{k}-\omega_{0})(t-\tau)],
\label{eq13}
\end{equation} 
is the delay Green's function. In fact, $G(t-\tau)$ is a two-time correlation function of the environment which measures the environment's memory effect. For the free-space case, the spectrum of the radiation field is infinitely broad and slowly varying, the memory effect is infinitesimally small and could be safely neglected. Then the Green function reduces to the form of Dirac delta function $G(t-\tau)=\beta_{0}\delta(t-\tau)$ which exhibits exponential spontaneous emission decay of the atomic excited-state population, where $\beta_{0}=\omega_{0}^{2}d^2/3\pi\varepsilon_{0}\hbar c^3$ is the decay rate \cite{scully97}. However, if the atom is embedded in a PCC, the decay rate is strongly modified, since the dispersion characteristic of radiation waves is deformed by the periodic dielectric structures of the PCC. In this paper, we consider two typical 
types of PCC that are widely investigated in previous literatures.

\subsection{Isotropic PCC}
Assuming the photonic density of the states becomes singular at the band-gap edge, the dispersion relation 
of the PCC near the band gap edge $\omega_{c}$ can be approximately written as \cite{john94}
\begin{equation}
\omega_{\textbf{k}}=\omega_{c}+A(k-k_{0})^2,
\label{eq14}
\end{equation}
where $\omega_{c}$ is the band edge frequency and $A\simeq\omega_{c}/k_{0}^{2}$ is a model dependent constant.
Note that this dispersion relation is isotropic (i.e., one-dimension) since it depends only on the magnitude $k$ of the wave vector. Starting from equation (\ref{eq14}) and using the Laplace transform method, we can obtain the expression of the amplitude \cite{john94,amri09}
\begin{equation}
q_{1}(t)=\frac{e^{x_{1}t}}{F'(x_1)}+\frac{e^{y_{1}t}}{H'(y_1)}+\frac{e^{i\delta t}}{\pi}\int_{0}^{\infty}\frac{\beta_{1}^{3/2}\sqrt{-iz}e^{-zt}}{i\beta_{1}^{3}-z(-z+i\delta)^2}dz, 
\label{eq15}
\end{equation}
where $\beta_{1}^{3/2}=\omega_{0}^{7/2}d^2/6\pi\varepsilon_{0}\hbar c^{3}$ and the detuning $\delta=\omega_{0}-\omega_{c}$. $x_{1}$ is the purely imaginary root of $F(x)=0$ in the region [$\text{Re}(x_{1})\leqslant0$ and $\text{Im}(x_{1})<\delta$], while $y_{1}$ is the complex root of $H(y)=0$ in the region [$\text{Re}(y_{1})<0$ and $\text{Im}(y_{1})<\delta$]. Here, the functions $F(x)$ and $H(y)$ are defined as $F(x)=x-i\beta_{1}^{3/2}/\sqrt{-ix-\delta}$ and $H(y)=y+\beta_{1}^{3/2}/\sqrt{iy+\delta}$, and $F'(x)=dF(x)/dx$ and $H'(y)=dH(y)/dy$.

\subsection{Anisotropic PCC}
Although the dispersion relation is very simple for the isotropic PCC, a real PCC in general has anisotropic structure in momentum space. Thus the photon dispersion relation dominated by the periodic dielectric structure is usually three-dimension. Numerical simulations show that the band edge is associated with a finite collection of symmetry related points $\textbf{k}=\textbf{k}_{0}^{i}$ rather than the isotropic case $|\textbf{k}|=k_{0}$. Then the effective anisotropic dispersion relation can be expressed approximately by \cite{quang97,zhu2000}
\begin{equation}
\omega_{\textbf{k}}=\omega_{c}+A|\textbf{k}-\textbf{k}_{0}^{i}|^2.
\label{eq16}
\end{equation}
Using this dispersion relation and following the similar calculations, we can obtain the expression of the amplitude $q_{3}(t)$ \cite{zhu2000,yang2000}
\begin{eqnarray}
q_{3}(t)=&&\frac{e^{x_{3}t}}{\mathcal{F}(x_3)}+\frac{e^{y_{3}t}}{\mathcal{H}(y_3)}+\frac{e^{i\delta t}}{\pi}\times\\
&&\int_{0}^{\infty}\frac{\beta_{3}^{3/2}\sqrt{iz}(\omega_{c}-iz)e^{-zt}}{i\beta_{3}^{3}z-[(\delta+iz)(\omega_{c}-iz)-\sqrt{\omega_{c}\beta_{3}^{3}}]^2}dz,\nonumber
\label{eq17}
\end{eqnarray}
where $\beta_{3}^{3/2}=(\omega_{0}d)^2\Sigma_{i}\sin^2\theta_{i}/8\pi\varepsilon_{0}\hbar A^{3/2}$, and $\theta_{i}$ is the angle between atomic dipole vector and $\textbf{k}_{0}^{i}$. The functions $\mathcal{F}(x)$ and $\mathcal{H}(y)$ are defined as $\mathcal{F}(x)=1-x^{2}/(2\beta^{3/2}\sqrt{-ix-\delta})$ and $\mathcal{H}(y)=1-iy^{2}/(2\beta^{3/2}\sqrt{iy+\delta})$.
$x_{3}$ is the purely imaginary root of $x-i\beta^{3/2}/(\omega_{c}+\sqrt{-ix-\delta})=0$ in the region [$\text{Re}(x_{3})\geqslant0$ or $\text{Im}(x_{3})>\delta$], while $y_{3}$ is the complex root of $y-i\beta^{3/2}/(\omega_{c}-i\sqrt{iy+\delta})=0$ in the region [$\text{Re}(y_{3})<0$ and $\text{Im}(y_{3})<\delta$].

Note that the existence of $x_{n}$ and $y_{n}$ with $n=1$, $3$, is highly dependent on the detuning $\delta$, i.e., the relative positions between atomic frequency $\omega_{0}$ and the band edge frequency $\omega_{c}$. The three terms in equations (\ref{eq15}) and (\ref{eq17}) determine three different emission fields, which are localized field, propagating field and diffusion field, respectively \cite{zhu1997}. However, unlike the isotropic case where $x_{1}$ and $y_{1}$ can exist together when the atomic transition frequency is far from the band edge and inside the gap, $x_{3}$ and $y_{3}$ can not coexist for the two-level atom in the anisotropic PCC \cite{zhu2000,yang2000}. This difference stems from the fact that the density of states is finite near the band edge in the anisotropic PCC rather than infinite in the isotropic case.
As we will show below, this difference will result in significant influence on the dynamics of spin squeezing.

Before we turn to discuss the dynamics of spin squeezing with atoms trapped in PCC, we remarkably note that the above dynamical procedures could be reformulated mathematically as a completely positive, trace-preserving (CPTP) linear map on the initial density operator of the qubit \cite{nielsen00}. The corresponding Kraus operators are given as
\begin{eqnarray}
E_{1}=\left(\begin{array}{cc}\sqrt{|q_{n}(t)|^2} & 0 \\0 & 1\end{array}\right),
E_{2}=\left(\begin{array}{cc}0 & 0 \\\sqrt{1-|q_{n}(t)|^2} & 0\end{array}\right)
\label{eq18}
\end{eqnarray}
where $n=1$ and $3$ denotes the isotropic and anisotropic PCC.
It is straightforward to check that
the Kraus operators of equation (\ref{eq18}) represent an amplitude damping noise. However, we emphasize that this amplitude damping noise should be non-Markovian since we have considered the memory effect of the PCC which has been registered self-consistently in the kernel function $G(t-\tau)$ in equation (\ref{eq12}).

\section{Robust spin squeezing preservation}
\label{sec:4}
According to the analysis of Refs. \cite{wang10,yin12}, calculating the dynamics of $N$-qubit spin squeezing can be reduced to determine the dynamics of two-qubit local expectations and correlations since the exchange symmetry of initial state always holds during the decoherence. Based on the Kraus operators given by equation (\ref{eq18}), we obtain the local expectations and correlations as follows
\begin{eqnarray}
\label{eq19}
&\langle\sigma_{z}^{(1)}\rangle=|q(t)|^2\langle\sigma_{z}^{(1)}\rangle_{0}+|q(t)|^2-1,\\
\label{eq20}
&\langle\sigma_{+}^{(1)}\sigma_{-}^{(2)}\rangle=|q(t)|^2\langle\sigma_{+}^{(1)}\sigma_{-}^{(2)}\rangle_{0},\\
\label{eq21}
&\langle\sigma_{-}^{(1)}\sigma_{-}^{(2)}\rangle=|q(t)|^2\langle\sigma_{-}^{(1)}\sigma_{-}^{(2)}\rangle_{0},
\end{eqnarray}
where $\langle \cdot\rangle_{0}$ is the expectation of initial state given by equations (\ref{eq6})-(\ref{eq8}). Substituting equations (\ref{eq19})-(\ref{eq21}) into equation (\ref{eq5}), we can eventually determine the explicit formula of the spin squeezing parameter $\xi_{R}^2$
\begin{equation}
\xi_{R}^2=\frac{1+2(N-1)|q(t)|^2\left[\langle\sigma_{+}^{(1)}\sigma_{-}^{(2)}\rangle_{0}-|\langle\sigma_{-}^{(1)}\sigma_{-}^{(2)}\rangle_{0}|\right]}{\left[|q(t)|^2\langle\sigma_{z}^{(1)}\rangle_{0}+|q(t)|^2-1\right]^2},
\end{equation}
as well as $\zeta_{R}^2=1-\xi_{R}^2$.
It is interesting to note that the spin squeezing parameter $\zeta_{R}^2$ is directly related to the time behavior of the single-qubit excited-state population $|q(t)|^2$. Therefore, one can control the spin squeezing by modulating the time-dependent function $|q(t)|^2$. As shown in equations (\ref{eq15}) and (\ref{eq17}), the behavior of $|q(t)|^2$ is mainly determined by the detuning $\delta=\omega_{0}-\omega_{c}$ and the dispersion relation of PCC. e.g., isotropic or anisotropic.

\subsection{Isotropic case}
\begin{figure}
  \includegraphics[width=0.45\textwidth]{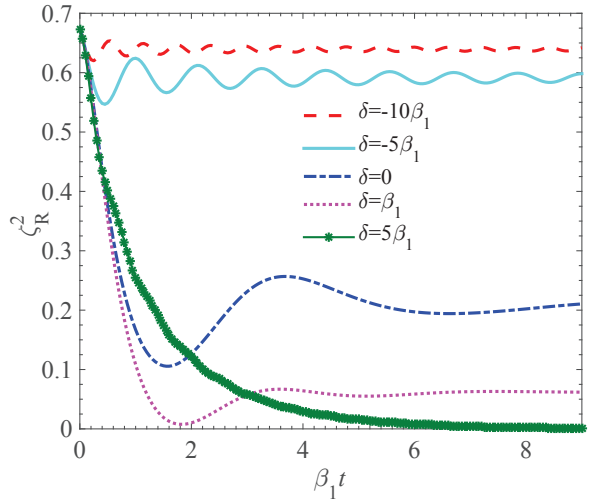}
\caption{(color online) Time evolution of the spin squeezing parameter $\zeta_{R}^2$ in the isotropic case, for various values of detuning. Dashed line $\delta=-10\beta_{1}$, solid line $\delta=-5\beta_{1}$, dotted-dashed line $\delta=-0$, dotted line $\delta=\beta_{1}$ and star-solid line $\delta=5\beta_{1}$. The other parameters are $\theta=0.15\pi$ and $N=10$. }
\label{Fig1}       
\end{figure}

Figure \ref{Fig1} shows the behavior of spin squeezing parameter $\zeta_{R}^2$ as a function of the scaled time $\beta_{1} t$ for various values of detuning under the isotropic dispersion. The dashed and solid lines show the results of spin squeezing when $\delta=-10\beta_{1}$ and $\delta=-5\beta_{1}$, namely, the atomic frequency is inside the band gap. We find that the spin squeezing shows rapid quasi-oscillations and finally yields to a definite value in the long-time limit. The quasi-oscillations and steady-state spin squeezing
can be all attributed to the formation of photon-atom dressed states \cite{zhu1997}. This can be understood as follows. As implied by equation (\ref{eq15}), the three radiation fields could coexist since the roots $x_{1}$ and $y_{1}$ can exist together when $\omega_{0}$ is near the band gap. The strong interactions between the atom and these radiation fields result in photon-atom dressed states. In this case, the interference between dressed states leads to the quasi-oscillatory behavior of the spin squeezing. Particularly, the bound dressed state with no decay is formed due to the permanent existence of localized filed. Therefore, a fractionalized steady-state spin squeezing is preserved even when $t\rightarrow\infty$. If the atomic frequency is tunneled far outside the band gap, the localized field disappears and no bound dressed state could be formed, then the spin squeezing decays exponentially (see star-solid line in figure~\ref{Fig1}).

\begin{figure}
  \includegraphics[width=0.45\textwidth]{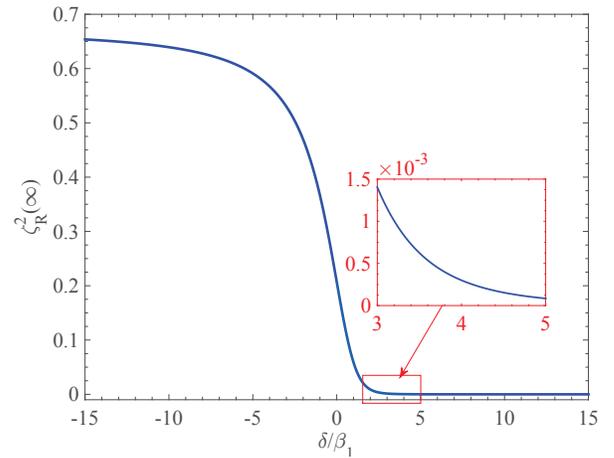}
\caption{(color online) Asymptotic value of spin squeezing $\zeta_{R}^2(\infty)$ as a function of $\delta/\beta_{1}$ in the isotropic case with the other parameters $\theta=0.15\pi$ and $N=10$. The inset clearly shows that the $\zeta_{R}^2(\infty)$ exponentially decreases but never equals to zero when the atomic frequency is outside the band edge.}
\label{Fig2}       
\end{figure}

To have a better understanding of the effect of detuning on spin squeezing preservation, we calculate the
steady-state spin squeezing. When time goes to infinity, the last two terms in equation (\ref{eq15}) turn to zero and only the first term remains, which contributes to the preservation of spin squeezing. Figure \ref{Fig2} plots the asymptotic value of spin squeezing parameter $\zeta_{R}^2(\infty)$ as a function of $\delta/\beta_{1}$.
We see that the steady-state spin squeezing is nearer to its maximum value for the atomic frequency farther from the band edge and deeper inside the gap. Remarkably, spin squeezing is partially preserved even when the atomic frequency lies outside of the band gap, but not far from the band edge since the bound state is still formed. As the atomic frequency is completely outside the band edge, the asymptotic value of spin squeezing exponentially decreases but never equals to zero, as displayed in the inset of figure \ref{Fig2}.

\subsection{Anisotropic case}

\begin{figure}
  \includegraphics[width=0.45\textwidth]{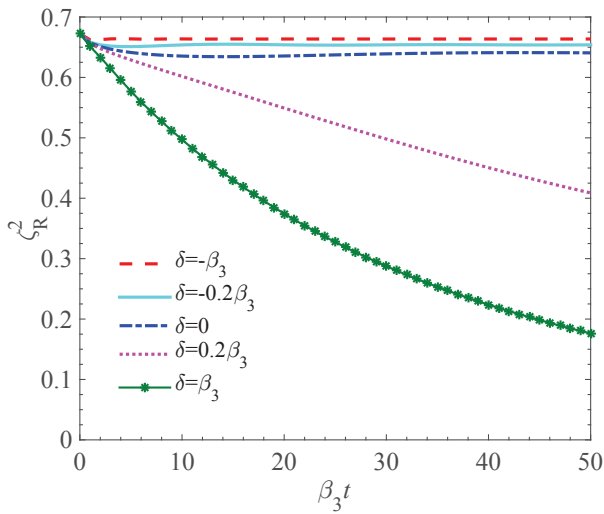}
\caption{(color online) Time evolution of the spin squeezing parameter $\zeta_{R}^2$ in the anisotropic case, for various values of detuning. Dashed line $\delta=-\beta_{3}$, solid line $\delta=-0.2\beta_{3}$, dotted-dashed line $\delta=-0$, dotted line $\delta=0.2\beta_{3}$ and star-solid line $\delta=\beta_{3}$. The other parameters are $\omega_{c}=100\beta_{3}$, $\theta=0.15\pi$ and $N=10$.}
\label{Fig3}       
\end{figure}

For the anisotropic PCC, as we mentioned above, the most significant difference is that $x_{3}$ and $y_{3}$ cannot coexist, namely, only one localized field or propagating field is present. Consequently, there is no interference between the dressed states. Hence, we expect the spin squeezing does not show quasi-oscillations during the evolution. Numerical simulation confirms this behavior, as shown in figure \ref{Fig3}.  When $\delta=-\beta_{3}$, $-0.2\beta_{3}$ or $0$, the bound state is formed due to the dressing of localized field, which results in the long-time preservation of spin squeezing, but no quasi-oscillations are observed since the propagating field is absent in this region. When $\delta=0.2\beta_{3}$ or $\beta_{3}$, the localized field disappears and only propagating field is present. As there is no bound state is formed, so the spin squeezing changes to zero.

\begin{figure}
  \includegraphics[width=0.45\textwidth]{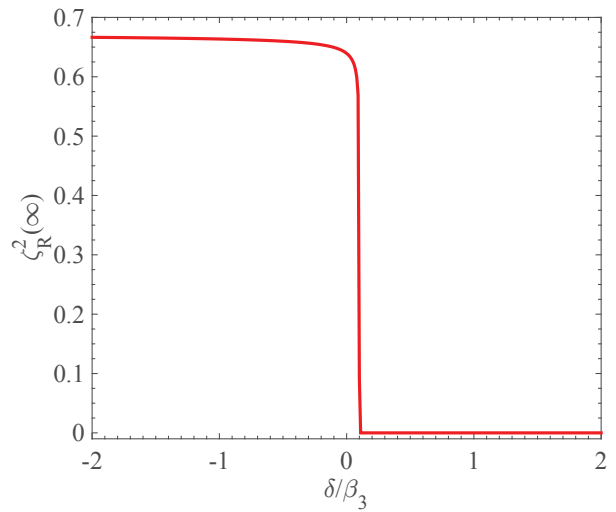}
\caption{(color online) Asymptotic value of spin squeezing $\zeta_{R}^2(\infty)$ as a function of $\delta/\beta_{3}$ in the anisotropic case with the other parameters $\omega_{c}=100\beta_{3}$, $\theta=0.15\pi$ and $N=10$.}
\label{Fig4}       
\end{figure}

Another manifestation of the anisotropic dispersion relation on the spin squeezing preservation is that the
asymptotic value of spin squeezing parameter $\zeta_{R}^2(\infty)$ becomes more sensitive to the changes of detuning than that in the isotropic case. Figure \ref{Fig4} clearly shows this difference. Unlike the result shown in figure \ref{Fig2}, where the asymptotic value of spin squeezing varies smoothly when the atomic frequency moves from the inside to the outside band edge, in the anisotropic case, $\zeta_{R}^2(\infty)$ experiences a sudden transition when $\omega_{0}$ is near the band edge $\omega_{c}$. The underlying physical reason is that the density of states has been assumed to be infinite near the band edge in the isotropic PCC. Due to the singularity of in density of states at $\omega_{c}$, the localized field is still existence when the atomic frequency is near $\omega_{c}$. Therefore, a fractionalized spin squeezing is preserved in the isotropic case. Even when the atomic frequency is completely outside the gap, the asymptotic value of spin squeezing only exponentially decreases but never equals to zero.
However, for the anisotropic case, the density of states is finite at $\omega_{c}$. The localized field suddenly disappears when $\omega_{0}$ approaches to the band edge from the inside gap, which leads to a sudden transition of $\zeta_{R}^2(\infty)$ when $\omega_{0}$ is near the band edge, for example, $\delta=0.1\beta_{3}$ when $\omega_{c}=100\beta_{3}$, as shown in figure \ref{Fig4}. In fact, if $\omega_{c}$ is large enough, the location of sudden transition point is close to $\delta=0$.

\section{Conclusions}
\label{sec:5}

To summarize, we have analyzed the spin squeezing dynamics of an ensemble of $N$ independent qubits coupled to their local and identical PCC. Thanks to the conserved exchange symmetry of qubits under decoherence, the exact expressions of spin squeezing parameter can be obtained by the two-qubit local expectations and correlations. Two typical types of PCC, i.e., isotropic and anisotropic PCC are considered.
We have shown that the spin squeezing can be drastically preserved in both two cases with the assistance of detuning modification. Particularly, there are quasi-oscillations of spin squeezing in the isotropic case, due to the interference of coexisted localized field and propagating field, while in the anisotropic case no quasi-oscillations occur. Intriguingly, it is noted that a sudden transition of steady-state spin squeezing emerges for the anisotropic case when the qubit frequency is near the band edge. This sudden transition is rooted in fact that the density of states of the anisotropic PCC is finite at the band edge.

We argue that the strategy presented in this paper, which enables the long time spin squeezing preservation against environmental noise, is potentially practical since neither complex reservoir engineering nor pulse series is required in our protocol. The robust spin squeezing preservation is attained only by utilizing a simple detuning modulation between the qubit and the band gap, which could be easily achieved in experiment, e.g., by Stark-shifting the qubit's frequency with a static electric field.
Our work is of great significance for quantum metrology in open systems and other quantum information processing tasks that the research objects are embedded in PCC.

\acknowledgments
This work is supported by the Funds of the National Natural Science Foundation of China under Grant No. 11247006 and No. 11365011, and by Scientic Research Foundation of Jiangxi Provincial Education Department under Grants No. GJJ150996.

\end{document}